\documentclass[showpacs,aps,twocolumn,prl]{revtex4}
\usepackage{graphicx}
\begin{document}
\bibliographystyle{apsrev}
\title{Direct Demonstration of the Anisotropic Origin of the
Macroturbulent Instability in Type-II Superconductors}
\author{L. M. Fisher}
\affiliation{All-Russian Electrical Engineering Institute, 12
Krasnokazarmennaya Street, 111250 Moscow, Russia}
\author{A. Bobyl, T. H. Johansen}
\affiliation{Department of Physics, University of Oslo, P.O. Box
1048, Blindern, 0316 Oslo 3, Norway}
\author{A. L. Rakhmanov}
\affiliation{Institute for Theoretical and Applied Electrodynamics
RAS,  13/19 Izhorskaya Street, 125412 Moscow, Russia}
\author{V. A. Yampol'skii}
\affiliation{Institute for Radiophysics and Electronics NANU,12
Proskura Street, 61085 Kharkov, Ukraine}
\author{A. V. Bondarenko, M. A. Obolenskii}
\affiliation{Kharkov National University, 4 Svoboda Sqr., 61077
Kharkov, Ukraine}

\begin{abstract}
The physical nature of the macroturbulence in the vortex matter in
YBCO superconductors is verified by means of magnetooptic study of
the instability in a single crystal prepared specially for this
purpose. The instability develops near those sample edges where
the oppositely directed flow of vortices and antivortices, guided
by twin boundaries, is characterized by the discontinuity of the
tangential component of the hydrodynamic velocity. This fact
directly indicates that the macroturbulence is analogous to the
instability of fluid flow at a surface of a tangential velocity
discontinuity in classical hydrodynamics, and is related to the
anisotropic flux motion in the superconductor.
\end{abstract}
 \maketitle

One of the most interesting and unexpected phenomena in  vortex
matter in type II superconductors has been discovered about ten
years ago~\cite{vl,ind,joh}. Using magneto-optical (MO) imaging
technique, the turbulent instability of the vortex-antivortex
interface has been observed in
$\mathrm{YBa_{2}Cu_{3}O_{7-\delta}}$ single crystals. When
magnetic flux is trapped in the superconductor and a moderate
field of opposite direction is subsequently applied, a boundary of
zero flux density will separate regions containing flux and
antiflux. In some temperature and magnetic field range such
flux-antiflux distribution can display unstable behavior
characterized by an irregular time-dependent propagation of the
boundary, where finger-like patterns often develop. This behavior
differs strongly from the predictions of the critical state model,
or creep models, where only quasi-static, or slow and regular
processes of flux redistribution can occur~\cite{bean,yes}.

The nature of this intriguing phenomenon remained unclear for a
long time. Actually, only a few attempts to interpret the
macroturbulent instability have been made. In Ref.~\cite{bass} the
problem was formulated in terms of the hydrodynamic flow of vortex
matter accompanied by a thermal wave generated by local release of
heat due to vortex-antivortex annihilation. However, as was
pointed out in Ref.~\cite{prl}, this mechanism is probably
irrelevant since the condensation energy within the vortex cores
is too small to give rise to a considerable overheating of the
superconductor.

It is essential that the macroturbulence is observed only in
$\mathrm{YBa_{2}Cu_{3}O_{7-\delta}}$ single crystals placed in a
magnetic field parallel to the crystallographic $c$-axis so that
the velocity vector of the moving magnetic flux lies in the
$ab$-plane. Specifically, the $\mathrm{YBa_{2}Cu_{3}O_{7-\delta}}$
material is characterized by a pronounced anisotropy of its
microstructure and of the physical properties in the $ab$-plane
due to the existence of twin boundaries. In particular, the twin
boundaries cause an anisotropic flow of the Abrikosov vortices
under the action of the Lorentz force. According to a number of
observations, the vortices move preferably along the twins, an
effect often referred to as flux guided motion~\cite{guid1,guid2}.
The examination of the vortex--antivortex flow under conditions of
the guiding effect prompted the authors of Ref.~\cite{prl} to
attribute the macroturbulent instability to the $ab$-plane
anisotropy. They assumed an analogy in the physical nature of the
macroturbulence with a kind of turbulence in the hydrodynamics of
usual fluids. According to a classical paper of Helmholtz,  the
flow of two fluids becomes unstable near their interface and
turbulence develops if there exists a discontinuity of the
tangential components of their velocities (see, for example,
Ref.~\cite{lan}). Such a tangential discontinuity is present at
the vortex-antivortex interface in twinned superconductors if the
twin boundaries are inclined at an angle $0<\theta < \pi/2$ with
respect to the direction of the Lorentz force. Indeed, the
anisotropy gives rise to vortex motion with a velocity component
normal to the Lorentz force. The vortices and antivortices are
forced to move towards each other along the interface where the
tangential component of the flux flow velocity is discontinuous.
Note that the role of anisotropy in the development of different
kinds of instabilities in superconductors was considered also by
Gurevich~\cite{G1,G2}. However, he did not study the problem of
the stability of the vortex-antivortex system.

With the above-mentioned physical picture in mind, a simple
hydrodynamic model was developed that takes into account the specific
features of vortex and antivortex motion in anisotropic
superconductors~\cite{prl}. To describe the anisotropic flux
motion, a linear relationship between the Lorentz force and the
vortex velocity with a symmetric tensor of viscosity was used. The
ratio $\varepsilon<1$ of the principal values of this tensor can
be used to characterize the anisotropy. The analysis of the
behavior of the vortex-antivortex system has shown that under
certain conditions the flat interface separating the regions
occupied by vortices and antivortices becomes unstable. The
instability develops at relatively small values of the anisotropy
parameter $\varepsilon<\varepsilon_c\ll 1$. It should be noted
that the parameter $\varepsilon$ describes the anisotropy of the
viscosity coefficient, and is in general different from the
critical current anisotropy of the superconductor. In order to
express $\varepsilon$ in terms of the current anisotropy one needs
to use the real current-voltage characteristics (CVC) of the
material. The linear relationship between vortex velocity and the
Lorentz force used in Ref.~\cite{prl} corresponds to having a
linear CVC. However, a nonlinear CVC should be used for a more
adequate approach to the problem.

The anisotropic power-law CVC in the form
\[
J_X=\frac{1}{\epsilon} J_c\left(\left|\frac{E_X}{E_0}\right|
\right)^{1/m}\mathrm{sign}\left(\frac{E_X}{E_0}\right),
\]
\begin{equation}\label{1}
J_Y=J_c\left(\left|\frac{E_Y}{E_0}\right|
\right)^{1/m}\mathrm{sign}\left(\frac{E_Y}{E_0}\right).
\end{equation}
was exploited in Ref.~\cite{jl} for the analysis of the
macroturbulent instability. Here $J_{X,Y}$ and $E_{X,Y}$ are the
current density and electric field components, $J_c$ is the
critical current density defined as the value of $J_Y$ at
$E_Y=E_0$ ($E_0=1 ~\mu$V/cm is the usual criterion), $\epsilon <1$
is the parameter of the current anisotropy, $X$ and $Y$ directions
correspond to those along and across the twin boundaries
(principal axes of the anisotropy). Typically, the parameter $m$
for YBCO single crystals is 10--20 or larger at temperatures
$T<50-60$~K. As it follows from Ref.~\cite{jl}, the instability
occurs if $m^2\epsilon^m < \varepsilon_c\ll 1$. As a result, the
macroturbulent instability can arise even for relatively low
\textit{current anisotropy}, $\epsilon \sim 0.3$--0.5. The model
developed in Refs.~\cite{prl,jl} allows one to describe the main
features of the macroturbulent instability. In particular, it
predicts a finite value of the wave number $k$ of the developed
perturbations and a temperature window in which the
macroturbulence can occur.

Thus, our previous studies~\cite{prl,jl} lead us to conclude that
the macroturbulent instability arises due to the tangential
discontinuity of the hydrodynamic velocity at the
vortex-antivortex interface resulting from the guiding effect.
Nevertheless, a certain dissatisfaction persisted since a direct
experimental confirmation of this physical picture has not been
obtained previously. In Ref.~\cite{joh} an attempt was made to
detect effects of the sample properties (such as sample structure,
size and geometry, current carrying capability) on the
macroturbulent behavior of the vortex matter. Unfortunately, this
study did not reveal direct correlations between the
macroturbulence and these properties. A subsequent experimental
study allowed us to conclude that the increase of the twin
boundary density results in an extension of the temperature window
in which the instability is observed~\cite{prl}. This result,
although being in favor of the anisotropic origin of the
macroturbulence, is insufficient as solid proof. This motivates
the present study devoted to a direct experimental demonstration
of the nature of the instability. The main idea of this paper is
to study the behavior of the flux-antiflux interface in a crystal
cut out in such a way that the anisotropy effects would be present
near some edges of the sample and absent near others. To realize
such an experiment, the sample was shaped into a triangular plate
with one edge cut parallel to the twin boundaries. Hence, flux
guiding and macroturbulence are not expected for the interface
running along this edge, but should be present for the other
edges.

The $\mathrm{YBa_{2}Cu_{3}O_{7-\delta}}$ single crystals were
grown using a technique described in Ref.~\cite{grown}. The
crystals were synthesized from CuO, Y$_2$O$_3$, and BaCO$_3$
powders of purity $99.99$ \%. Powders containing the metallic
elements Cu:Ba:Y in the ratio 73:24.5:1.5 were mixed and annealed
in flowing oxygen at 1130~K for 4 days. The crystals were grown in
a gold crucible, in the temperature range of 1130--1250~K, in the
presence of a temperature gradient of 2--4~K/cm with the rate of
temperature decrease of about 4 K/hour. This method allows us to
produce crystals with dimensions up to $5\times 5$~mm$^2$ parallel
to the $ab$ plane and about 10--20~$\mu$m along the $c$ axis. The
crystals were saturated with oxygen at a temperature of 700~K in
an oxygen flow at ambient pressure for four days. Then, several
crystals having large domains with aligned twin boundaries were
chosen. After a selection of such domains, we prepared two samples
and shaped them by laser cutting~\cite{cut} into a nearly
right-angled triangular plate. The polarized light microscope
image of one of the samples is shown in Fig.~\ref{f1}. The size of
the sample along the hypotenuse is about 1.1~mm. The
crystallographic $ab$-plane coincides with the sample plane. It is
clearly seen that the twin boundaries are directed along the
hypotenuse. The twin spacing is approximately  2 $\mu$m. The
critical temperature of the samples is 91~K and the width of the
transition is about 0.3~K.

The study of the magnetic flux penetration and the macroturbulence
was carried out by the conventional magnetooptic imaging technique
\cite{Dorosinskii92}. The image in Fig.~\ref{f2} demonstrates the
distribution of trapped magnetic flux after cooling the sample
down to 30~K in an external transverse magnetic field $H=1$~kOe,
which was subsequently switched off. The brighter regions of the
image correspond to higher values of the magnetic induction. The
anisotropy of the field distribution is clearly seen, and one can
evaluate the critical current density and its anisotropy using
this image. The evaluation gives $J_c$ about $10^5$~A/cm$^2$ for
the critical current density along the twin boundaries. The
anisotropy parameter (the ratio of the critical current densities
across and along twins) $\epsilon$ can be estimated using the
geometrical construction shown in Fig.~\ref{f2}. According to the
critical state model and current conservation law, one has
$$
\epsilon = \frac{\sin\alpha}{\sin \beta} = \frac{OB}{OA}\ ,
$$
which amounts to 0.35 using the values of $\alpha$ and $\beta$
seen in the figure. Thus, this sample is suitable for testing the
nature of the macroturbulent instability.

In order to search for macroturbulence, the sample was first
cooled in a transverse magnetic field $H$. Then the field was
abruptly reversed and MO images were recorded and analyzed.
Various sample temperatures and reverse fields were used. The most
pronounced unstable behavior was observed at $H=1$~kOe and
$T=30$~K, and is illustrated by the series of MO images in
Fig.~\ref{f3}. The manifestation of instability as seen through
the oculars of the microscope can be described as follows. At
first, a small-scale and fast 'trembling' of the interface between
flux and antiflux (the dark lines in the images) was observed. It
is clearly seen from the images that the magnetic flux frozen in
the central part of the sample disappears with time.
Unfortunately, we are not able to illustrate this effect by a
static photographs. Then, a bending and irregular motion of the
interface deep into the sample occurred. However, we could not
observe a very distinct fingering of the interface, as found in
the  classical observations~\cite{vl,ind,joh} of the phenomenon.
On the other hand, previous studies~\cite{unpub} have shown that
lack of fingering is typical when the lateral dimensions of the
sample are comparable to the spatial scale of the turbulent
perturbations. Unfortunately, we could not at present produce
larger samples with desirable geometry and with a single twin
boundary orientation.

The images (a-d) in Fig.~\ref{f3}, obtained at 0.1, 0.2, 0.3 and
10 seconds after the field reversal, show the consecutive stages
of the development of the instability. The key point here is to
observe the significant difference in the interface geometry, and
its motion away from the sample edges as function of time. First,
one notes that except for the hypotenuse, the interface elsewhere
is very sharply defined, a usual feature of turbulent behavior.
Second, along the hypotenuse the interface remains essentially
static whereas substantial motion takes place elsewhere, e.g. for
the interface along the upper cathetus, where the velocity is
estimated to 3~mm/sec at the initial stage of the development of
the instability. Note that the fast change of interface position
occurs after the field reversal, when the critical profile has
been established, and can be interpreted as the development of the
instability only. Unstable motion appears clearly also from the
short edge in the lower left part of the crystal. Also along the
left cathetus the interface moves, although with a slightly
smaller velocity. We conclude therefore that we experimentally
have found that macroturbulent instability occurs only along edges
oriented with some angle $\theta \ne 0, \pi/2$ with respect to the
twin boundaries, i.e., where the guiding effect leads to the
tangential discontinuity of the hydrodynamic vortex velocity. By
performing MO imaging at different temperatures we found that the
macroturbulent instability exists in the present sample in the
temperature window of 15~K$<T<45$~K. The effect is well
reproducible after several cycling magnetic field and temperature.

Thus, the present study can be considered as a crucial experiment
for the ascertainment of the physical nature of the macroturbulent
instability in type-II superconductors. The specific anisotropy
for $\mathrm{YBa_{2}Cu_{3}O_{7-\delta}}$ superconductors provides
the guiding effect in the vortex motion. As a result, the
discontinuity of the tangential component of the flux-line
velocities appear at the vortex-antivortex interface. This leads
to the development of the turbulence similar to the case of the
classical dynamics of fluids.

We acknowledge C.J. van der Beek for preliminary measurements.
This work is supported by INTAS (grant 02--2282), RFBR (grants
03-02-17169a, 03-02-16626a), Russian National Programm on
Superconductivity (contract~40.012.1.1.11.46), and the Research
Council of Norway.

\newpage
\begin{figure}[tbp]
\caption{\label{f1} Polarized light image of the sample, which
reveals that it consists of essentially singly oriented twin
boundaries parallel to the long side (hypotenuse).}
\end{figure}

\begin{figure}[tbp]
\caption{\label{f2} MO image of the trapped magnetic flux.
$T=30$~K.}
\end{figure}
\begin{figure}[tbp]
\caption{\label{f3} Evolution of the magnetic flux distribution
under condition of the development of instability. Images (a)-(d)
were obtained in 0.1, 0.2, 0.3, 10 seconds after the reverse of
the external magnetic field, respectively. $T=30$~K, $H=1$~kOe.
The flux-antiflux interface is the dark line, e.g. as pointed at
by the white arrow.}
\end{figure}
\end{document}